\documentclass[aps,twocolumn]{revtex4}
\usepackage{amssymb}
\usepackage{amsfonts}
\usepackage{amsmath}
\usepackage{graphicx}

\newcommand{\beq}{\begin{equation}}
\newcommand{\eeq}{\end{equation}}

\begin{document}

\title{Transient entanglement in a spin chain stimulated by phase pulses.}
\author{Isabel Sainz$^{1}$,Gennadiy Burlak$^{2}$, and Andrei B. Klimov$^{1}$$%
\thanks{%
klimov@cencar.udg.mx}$}
\affiliation{$^{1}$Departamento de F\'{\i}sica, Universidad de Guadalajara, Revoluci\'{o}%
n 1500, Guadalajara, Jalisco, 44420, M\'{e}xico.}
\affiliation{$^{2}$ Centro de Investigaci\'{o}n en Ingenier\'{\i}a y Ciencias Aplicadas,
Universidad Aut\'{o}noma del Estado de Morelos, Cuernavaca, Morelos, M\'{e}%
xico}
\pacs{PACS number}

\begin{abstract}
Dynamics of the one-dimensional open Ising chain under influence of $\pi $%
-pulses is studied. It is shown that the application of a specific
sequence of such instant kicks to selective spins stimulates arising of
perfect dynamical pairwise entanglement between ends of the spin chain.
Analytic formulas for the concurrence dynamics are derived.
It is also shown that the time
required to perfectly entangle the ends of the chains grows linearly with the
number of spins in the chain. The final entangled state of the ending
spins is always the same and does not depend on length the chain.
\end{abstract}

\maketitle

\section{Introduction}

Generation and distribution of entanglement between remote particles is a
necessary requirement for quantum information transfer \cite{transfer1},
\cite{transfer2}. It turns out that spin chains are one of the most
promising quantum channels for creation of spin-spin entanglement \cite%
{RMP08}. It appears, that in spite of existence of correlations in spin
chains, the amount of entanglement generated between two particular spins
rapidly decays with the distance between them \cite{RMP08}, \cite{bose CP}
even near the point of quantum phase transition \cite{qpt}. So that, the
implementation of efficient mechanisms for entangling spatially separated
subsystems is still a challenging problem, especially for massive particles.

During the last years different types of spin-spin interactions have been
extensively studied as possible candidates to entangle remote particles. It
has been proved that for certain types of spin-$1/2$ open Heisenberg chains,
it is possible to entangle the ends of the chain by action of stationary
Hamiltonians \cite{bosePRL} and implementing different mechanisms like
mirror symmetry \cite{mirror}, dimerized models or weak couplings at the
ends of the chain \cite{dimer}, performing a quench \cite{quench1}, \cite%
{quench2}, or considering alternating couplings between the spins \cite{nest}%
.\textbf{\ }Also,\textbf{\ }several dynamical schemes, such as application
of optimized time-dependent magnetic field \cite{boseTD} and periodically
time-dependent nearest-neighbor coupling \cite{galve} have been proposed
to faithfully entangle the ends of an open spin chain.
Unfortunately, only a partial (although a quite large) entanglement can be
achieved with these schemes. Besides, owing mathematical complexity of the
problem, the main tool for studying such systems remains numerical
simulations.

The simplest type of interaction between massive particles is the
homogeneous Ising chain. Being the simplest type of interaction, it has been
a candidate for implementation of quantum information algorithms since the
very beginning \cite{nuclear1}. It worth noting a recent proposal for
measurement-based quantum computation \cite{nuclear2} using homogeneous
Ising-like interaction. Also, this kind of interaction accompanied by the
global $\pi $ pulses allows \ a perfect qubit transport, quantum mirrors and
universal quantum computation \cite{twamley}.

On the other hand, the possibility of application of pulses to quantum
systems in order to preserve the coherence was discussed several years ago
\cite{equidistant}, and recently, the idea of applying optimized $\pi $ pulses at
irregular intervals of time was proposed and studied by Uhrig \cite{Uhrig}
for efficient control of dephasing in spin systems and later applied for
preserving entanglement in dephasing environments \cite{agarwal}.

In this article we propose a simple \textit{analytical }scheme which allows
to create a transient entanglement between the ends of an homogeneous spin
chain by means of application of a sequence of $\pi $ pulses on some
particular spins. We apply $\pi $ pulses not to all qubits (as it was
proposed in \cite{twamley} to control a qubit transport) but only to a part
of them, which results in a perfect transient entanglement between the ends
the chain.

\section{The model}

Let us consider an open chain of $N$ spins $1/2$ with homogeneous Ising-like
interaction (the coupling constant is taken to be unity), governed by the
Hamiltonian
\begin{equation}
H=\sum_{j=1}^{N}\hat{s}_{zj}\hat{s}_{zj+1},  \label{Hint}
\end{equation}%
and initially prepared in the coherent superposition
\begin{equation*}
\left\vert \Psi _{0}\right\rangle =\Pi _{j=1}^{N}\left\vert +\right\rangle
_{j},\quad \left\vert +\right\rangle _{j}=\frac{1}{\sqrt{2}}\left(
\left\vert 0\right\rangle _{j}+\left\vert 1\right\rangle _{j}\right) .
\end{equation*}%
Since $\left\vert \Psi _{0}\right\rangle $ is not an eigenstate of the
Hamiltonian Eq.(\ref{Hint}), some specific spin-spin correlations arise
during the Hamiltonian evolution, $\hat{U}(t)=e^{-itH}$. The evolution of
the density matrix $\hat{\rho}$ can be found in the following closed form,
\begin{eqnarray}
\hat{\rho}(t) &=&\hat{U}(t)\left\vert \Psi _{0}\right\rangle \left\langle
\Psi _{0}\right\vert \hat{U}^{\dag }(t)  \notag \\
&=&\left( \frac{1}{2}+\cos \frac{t}{2}\hat{s}_{x1}+2\sin \frac{t}{2}\hat{s}%
_{y1}\hat{s}_{z2}\right)   \notag \\
&&\prod_{j=2}^{N-1}\left[ \frac{1}{2}+\cos ^{2}\frac{t}{2}\hat{s}_{xj}-4\sin
^{2}\frac{t}{2}\hat{s}_{xj}\hat{s}_{zj-1}\hat{s}_{zj+1}\right.   \notag \\
&&\left. +2\sin \frac{t}{2}\cos \frac{t}{2}\hat{s}_{yj}\left( \hat{s}_{zj-1}+%
\hat{s}_{zj+1}\right) \right]   \notag \\
&&\left( \frac{1}{2}+\cos \frac{t}{2}\hat{s}_{xN}+2\sin \frac{t}{2}\hat{s}%
_{yN}\hat{s}_{zN-1}\right) ,  \label{stateNK}
\end{eqnarray}%
In the Hamiltonian (\ref{Hint}) only nearest neighbors interact, and a
dynamical pairwise entanglement is generated between them. As it can easily
be seen from (\ref{stateNK}), the entanglement dynamics between all the
pairs in the middle of the chain is the same, the concurrence \cite%
{Wootters:1998a}, is given by
\begin{equation}
C_{jj+1}=\max \left[ 0,\frac{|\sin t|}{2}-\frac{\sin ^{2}\frac{t}{2}}{2}%
\right] ,  \label{C_middle}
\end{equation}%
where $j=2,\dots ,N-2$ and its maximum is $\sim 0.31$. Since the spin chain
is open, the concurrences for the spins $1$ and $2$, and the spins $N-1$ and
$N$ are different that given in Eq. (\ref{C_middle}),
\begin{equation}
C_{12}=C_{N-1N}=\frac{|\sin t|}{2},  \label{C_ini}
\end{equation}%
reaching the maximum of $1/2$ at $t=(2n+1)\pi /2,$ with $n=0,1,\ldots $.

It is important to stress that non neighboring pairs of spins will never get
entangled under the Hamiltonian evolution.

\section{Entangling the ends of the chain with $\protect\pi$-pulses}

In order to generate entanglement between the ends of the chain let us apply
instant $\pi $ pulses to the first $N-1$ spins at certain times $t_{j}$.
Such kicks correspond to rotations around the $y$-axis, and for the $j$-th
spin are represented by the operator $\hat{R}_{j}=e^{-i\pi \hat{s}_{yj}/2}$.
The main idea is as follows: initially all the spins are in eigenstates of $%
\hat{s}_{xj}$ so that, in the course of the evolution governed by Eq.(\ref%
{Hint}), some spins become correlated. At the instant $t_{1}=\pi $ we apply
a $\pi $-rotation to the first $N-1$ spins, producing flip of those spins.
Afterwards, the evolution under the Hamiltonian Eq.(\ref{Hint}) allowed to
continue leading to the recoupling of all the spins. Then, at the instant $%
t_{2}=2\pi $ we apply the inverse $\pi $-rotation, allowing the system to
evolve further and repeat this cycle until $N-2$ spin flips are performed at
times $t_{j}=j\pi $, $j=1,2,...,N-2$, so that
\begin{eqnarray*}
\left\vert \Psi (t)\right\rangle &=&\hat{U}(t-\pi (N-2))\hat{R}^{(-1)^{N-1}}
\hat{U}(\pi )\ldots \\
&&\ldots\hat{U}(\pi )\hat{R}^{-1}\hat{U}(\pi )\hat{R}\hat{U}(\pi )\left\vert
\Psi _{0}\right\rangle ,  \label{psi_t}
\end{eqnarray*}
where $\hat{R}=\otimes \Pi _{j=1}^{N-1}\hat{R}_{j}$. Using (\ref{psi_t}) we
now proceed to compute the reduced density matrix and the (time-dependent)
concurrence $C(t)$ \cite{Wootters:1998a}.

The three-spin case is special, since the middle spin is coupled to both
ends. For a three-spin chain a single $\pi $ pulse at $t_{1}=\pi $, applied
to the first and second spins, produces entanglement between the ends. It is
straightforward to obtain the concurrence for the first and third spins:
\begin{equation*}
C_{13}(t)=\left\{
\begin{array}{cc}
\cos^{2}\frac{t}{2}, & t\geq \pi \\
0, & t<\pi%
\end{array}%
\right. ,
\end{equation*}%
i.e. the first and third spins, which were not entangled before the pulse ($%
t<\pi $), become completely entangled at $t=2\pi k,~k=1,2,\ldots $, while $%
C_{12}(t)=C_{23}(t)=0$ for $t\geq \pi $.

For longer chains the dynamical behavior of the pairwise entanglement is
rather unusual. After the first kick and before the $(N-2)$-nd kick the
pairwise entanglement between any pair of spins disappears. Since the system
is closed, the pairwise entanglement transforms into a multipartite
entanglement during that time. Moreover, after the $\left( N-2\right) -$nd
kick it is possible to obtain the following analytic expression for the
concurrence between the ends of the chain:

a) for the even number of spins
\begin{equation}
C(t)=\left\{
\begin{array}{cc}
\max\left[0,|\sin \frac{t}{2}|-\frac{\cos^2\frac{t}{2}}{2}\right], &
t\geq(N-2)\pi \\
0, & t<(N-2)\pi%
\end{array}
\right. ,  \label{C_even}
\end{equation}
and

b) for the odd number of spins
\begin{equation}
C(t)=\left\{
\begin{array}{cc}
\max\left[0,|\cos \frac{t}{2}|-\frac{\sin^2\frac{t}{2}}{2}\right], &
t\geq(N-2)\pi \\
0, & t<(N-2)\pi%
\end{array}
\right.,  \label{C_odd}
\end{equation}
while the nearest neighbors in the middle of the chain recover their
original transient entanglement $C_{jj+1}(t)=\max [0,|\sin t|-\sin
^{2}t/2]/2 $ for times $t>(N-2)\pi $.

Thus, the subsystem containing the first and the last spins of the chain
stays in an unentangled state until the last kick is applied. As a result
the subsystem reaches a maximally entangled state at the moments $t=\left(
N+2k-1\right) \pi ,~k=0,1,\ldots $ , being unentangled with the rest of the
chain, in accordance with the monogamy property \cite{monogamy}\textbf{. }So
that, the time needed to completely entangle the ends of the chain linearly
depends on the length of the chain. In particular, at the instant $%
t=(N-1)\pi $ the state of the chain is as follows:
\begin{eqnarray*}
\left\vert \Psi \right\rangle  &=&\frac{1}{2^{(N-1)/2}}\prod_{j=2}^{N-1}%
\left[ \left\vert 1\right\rangle _{j}+(-1)^{j}\left\vert 0\right\rangle _{j}%
\right]  \\
&&\left[ \left\vert 00\right\rangle _{1N}+\left\vert 11\right\rangle
_{1N}+i\left( \left\vert 01\right\rangle _{1N}+\left\vert 10\right\rangle
_{1N}\right) \right] .
\end{eqnarray*}

On the other hand, the concurrence between the spins 1 and 2, and the spins $%
N-1$ and $N$, remains zero for any $t>\pi $.

\begin{figure}[tbp]
\begin{center}
\includegraphics[width=0.5\textwidth]{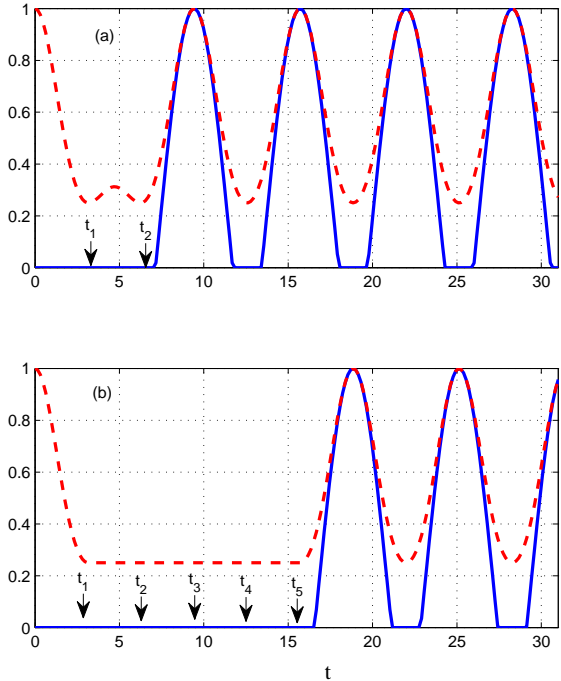}
\end{center}
\caption{(Color online) Dynamics of concurrence $C(t)$ for the chain with
(a) $4$ spins for $t_{1}=\protect\pi $, $t_{2}=2\protect\pi $; and (b) $7$
spins for $t_{1}=\protect\pi $, $t_{2}=2\protect\pi ,t_{3}=3\protect\pi $, $%
t_{4}=4\protect\pi $, $t_{5}=5\protect\pi $. Dashed line shows $Tr(\protect%
\rho _{1N}^{2})$, arrows indicate when the kicked pulsed are applpied.}
\label{Pic_C_4_7}
\end{figure}

In Fig.(\ref{Pic_C_4_7}) we show the dynamics of the concurrence $C(t)$ for
the chain with (a) $4$ spins and $\pi $-pulses applied to the first $3$
spins at $t_{j}=j\pi $, $j=1,2$, and (b) $7$ spins and $\pi $-pulses applied
to the first 6 spins at $t_{j}=j\pi $, $j=1,...5$. The dashed line shows the
purity of the of subsystem composed by the first and the last spins of the
corresponding chain, $Tr(\rho _{1N}^{2})$. Observe that, due to the
Hamiltonian evolution, the concurrence and purity keeps oscillating after
application of the last kick.

\begin{figure}[ptb]
\begin{center}
\includegraphics[width=0.5\textwidth]
{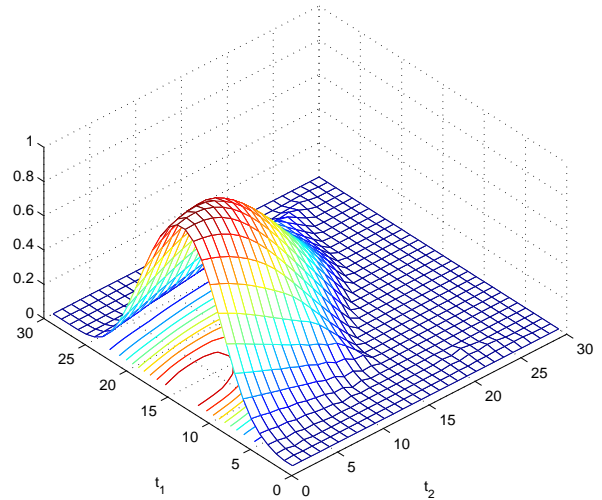}
\end{center}
\caption{(Color online) The concurrence $C$ \ for chain of $4$ spins as a
function of the moments of applied kicks at fixed time $t=3\protect\pi$.}
\label{Pic_C_t1_t2}
\end{figure}

In order to study the behavior of concurrence $C(t)$ when the times $t_{j}$
of application of the $\pi $-pulses are not commensurable, we have
numerically calculated $C(t)=C(t,t_{1},t_{2})$ as a function of $t_{1}$ and $%
t_{2}>t_{1}$ at different fixed times. In Fig.\ref{Pic_C_t1_t2} we plot the
concurrence $C(t)$ for $4$ spins at time $t=3\pi $ (compare with Fig.\ref%
{Pic_C_4_7} (a)) when $t_{1}$ $\in \lbrack 0.1,5]$ and $t_{2}\in \lbrack
5.1,9]$ . We can observe that $C(t_{1},t_{2})$ has a pronounced maximum, $%
C=1,$ at $t_{1}=\pi $, and $t_{2}=2\pi $ and smoothly decreases when $t_{1,2}$
deviate from these optimal values.

\section{Discussion}

We have studied the dynamics a one-dimensional open Ising chain of $N$ spins
assisted by phase kicks at times $t_{j}=j\pi $, $j=1,2,...,N-2$. It is found
that the application of $N-2$ instant pulses to $N-1$ spins leads to arising
of transient perfect entanglement between the first and the last spins of
the chain. One interesting results is that, if the number of pulses is less
than required, then every pairwise concurrence would be zero. The periodic
behavior of the concurrence between the ends of the chain, $C_{1N}(t)$
reaches the maximum value $C=1$ at some specific times even when the pulses
are stopped being applied and the chain evolves under the Hamiltonian (\ref%
{Hint}) only. This effect substantially enlarges possible applications of
the low-dimensional spin chains in quantum information technology.

On the other hand, in this particular model we may observe\textbf{\ }an
important example of a non-trivial effect of local transformations on the
entanglement production in non-linear systems.

It is easy to see that combining the present scheme with quench one can
entangle any two spins $r$ and $s$, $1<r<s<N$, in the chain, i.e. to model a
quantum router. Really, we just have to \textquotedblleft
disconnect\textquotedblright\ the chain \cite{quench2} from 1 to $r-1$ and from $s+1$ to $N$
 and then apply the above discussed sequence of pulses.

It worth noting that similar local transformations in the form of instant
pulses were also used for another purposes: for atomic squeezing enhancement
in the Dicke states \cite{daisuke}, for intensification of the entanglement
in continuous-variables \cite{cv} and in two-spin \cite{PRA} systems.

This work is partially supported by the Grant 106525 of CONACyT (Mexico)

\end{document}